\documentclass[rapids]{jfm}

\usepackage{natbib}
\usepackage{graphicx}
\usepackage{bm}
\usepackage{xcolor}
\usepackage{makecell}
\usepackage{amsmath}
\usepackage{amsfonts}
\usepackage{hyperref}
\hypersetup{
    colorlinks = true,
    urlcolor   = blue,
    citecolor  = black,
}

\newcommand{\RomanNumeralCaps}[1]
\definecolor{myrd}{RGB}{227,26,28}
\definecolor{myrl}{RGB}{251,154,153}
\definecolor{myod}{RGB}{255,127,0}
\definecolor{myol}{RGB}{253,191,111}
\definecolor{mygd}{RGB}{51,160,44}
\definecolor{mygl}{RGB}{178,223,138}
\definecolor{mybd}{RGB}{31,120,180}
\definecolor{mybl}{RGB}{166,206,227}
\definecolor{mypd}{RGB}{106,61,154}
\definecolor{mypl}{RGB}{202,178,214}
\definecolor{mygr}{RGB}{255,255,153}
\newcommand{\vect}[1]{\boldsymbol{#1}}
\newcommand\NavSto{Nav\-ier--Sto\-kes}

\newcommand\threedity{three-di\-men\-sion\-al\-ity}
\newcommand\twod{two-di\-men\-sion\-al}
\newcommand\twodity{two-di\-men\-sion\-al\-ity}
\newcommand\qtwod{quasi-two-di\-men\-sion\-al}

\newcommand\TS{TS}
\newcommand\ie{i.e.}
\newcommand\ii{\mathrm{i}}

\newcommand\Fig{Figure} % Start of sentence FOR JFM or PRF
\newcommand\fig{figure}  % Mid sentence FOR JFM
\newcommand\alphaOpt{\alpha_\mathrm{opt}}

\newcommand\tauOpt{\tau_\mathrm{opt}}

\newcommand\ELD{E_\mathrm{D}}
\newcommand\EB{E_\mathrm{B}}

\newcommand\Ezero{E_0}
\newcommand\Euv{E}

\newcommand\rrc{r_\mathrm{c}}
\newcommand\ReyCrit{\Rey_\mathrm{c}}

\newcommand\Np{N_\mathrm{p}}
\newcommand\Nf{N_\mathrm{f}}
\newcommand\amax{\alpha_\mathrm{max}}

\newcommand\meanfoco{\bar{c}_\kappa}

\newcommand\bnp{\nabla_\perp}
\newcommand\bnabp{\bnabla\!\!_\perp}

\newcommand\pp{p_\perp}

\newcommand\Uvp{\vect{U}_\perp}
\newcommand\uvp{\vect{u}_\perp}
\newcommand\vp{v_\perp}
\newcommand\vtp{\tilde{v}_\perp}
\newcommand\up{u_\perp}

\newcommand\vhp{\hat{v}_\perp}
\newcommand\uvph{\hat{\vect{u}}_\perp}

\newcommand\vhpn{\hat{v}_{\perp,n}}

\newcommand\vhpnnm{\hat{v}_{\perp,n,\lvert n\rvert+2m}}

\newcommand\woo{\hat{v}_{\perp,1,1}}
\newcommand\wno{\hat{v}_{\perp,-1,1}}
\newcommand\wtt{\hat{v}_{\perp,2,2}}

\newcommand\uot{\hat{u}_{\perp,0,2}}

\newcommand\xtp{\tilde{\xi}_\perp}

\DeclareMathOperator{\sign}{sgn}

\title{Subcritical transition to turbulence in \qtwod\ shear flows}

\author{Christopher J. Camobreco\aff{1,2},
        Alban Poth{\'e}rat\aff{3}\corresp{\email{alban.potherat@coventry.ac.uk}}
 \and Gregory J. Sheard\aff{1}}

\affiliation{\aff{1}Department of Mechanical and Aerospace Engineering, Monash University, VIC 3800, Australia
             \aff{2}Department of Mechanical Engineering, University of Melbourne, VIC 3010, Australia
             \aff{3}Fluid and Complex Systems Research Centre, Coventry University, Coventry CV15FB, United Kingdom}

\begin{document}
\maketitle

\begin{abstract}
The transition to turbulence in conduits is among the longest-standing problems in fluid mechanics.
Challenges in producing or saving energy hinge on understanding promotion or suppression of turbulence. While a global picture based on an intrinsically 3D subcritical mechanism is emerging for 3D turbulence, subcritical turbulence is yet to even be observed when flows approach two dimensions, \eg under intense rotation or magnetic fields.
Here, stability analysis and direct numerical simulations demonstrate a subcritical quasi-2D transition from laminar flow to turbulence, via a radically different 2D mechanism to the 3D case, driven by nonlinear Tollmien--Schlichting waves.
This alternative scenario calls for a new line of thought on the transition to turbulence and should inspire new strategies to control transition in rotating devices and nuclear fusion reactor blankets.
\end{abstract}


\section{Introduction}
\label{sec:intro}
One of the most important questions in fluid mechanics is how, and under what conditions, flows transition to a turbulent state; this determines the topological, dissipative and mixing properties of these flows. Besides its fundamental interest as a unique physical process, it is central to every application where fluid flows through a conduit: turbulent mixing promotes heat exchange in cooling applications, whereas turbulent dissipation drastically increases energy consumption. As discovered by \citet{Reynolds1883experimental}, the flow of water through a pipe became turbulent only for sufficiently high values of the eponymous Reynolds number $Re=U_0 L/\nu$. In the present work, $Re$ is built out of the typical streamwise velocity $U_0$ and kinematic viscosity $\nu$ of the fluid, and the transverse conduit length scale $2L$. Since then, the question of transition was often tackled by seeking the conditions necessary for perturbations growing from the laminar base flow to ignite turbulence.

In pipes and other shear flows, the most distinctive property of the transition to turbulence is that it is subcritical. If $Re_c$ is the critical Reynolds number beyond which some perturbation grows exponentially from an infinitely small amplitude through a linear mechanism, turbulence can develop at $Re<Re_c$, provided the flow is seeded with a sufficiently energised perturbation. The process is nonlinear and amplifies finite amplitude perturbations through an intrinsically 3D `lift-up' mechanism, enacted by the growth of streamwise streaks \citep{Schmid2001stability}.
It also becomes active at Reynolds numbers well below $Re_c$, where infinitesimal perturbations are severely damped. For convenience we define $\rrc=Re/Re_c$, where subcritical Reynolds numbers correspond to $\rrc<1$. The least damped perturbations are the 2D transverse-invariant \emph{Tollmien--Schlichting waves} (\TS\ waves), which manifest in plane shear flows. For instance, \cite{Beneitez2019edge} showed that \TS\ waves were not found to partake in the 3D transition to turbulence below $\rrc \simeq 0.98$ in Blasius boundary layers, while \cite{Zammert2019transition} showed that they could not be detected in plane Poiseuille flow for $r_c \lesssim 0.84$.

However, in rapidly rotating or stratified flows, or in an electrically conducting fluid subjected to a high magnetic field, fluid motion can be prevented from becoming 3D if the respective Coriolis, buoyancy or Lorentz forces are sufficiently intense.
Hydraulic circuits in rotating machines, atmospheres, oceans and some models of planetary interiors subject to planetary rotation, and the liquid metal blankets cooling fusion reactors, all occupy this category. Real flows can never be fully 2D \citep{Paret1997electromagnetically, Akkermans2008amplification}; \threedity\ subsists either in asymptotically small measure or in asymptotically small regions such as boundary layers \citep{Zavala2000nonlinear, Potherat2012three}.
The resulting flows are \emph{quasi-2D}. Since the lift-up mechanism driving transition in 3D shear flows cannot manifest in quasi-2D flows, can subcritical quasi-2D turbulence exist, and if so via which alternative transition mechanism?

Traditionally, quasi-2D turbulence has ``only'' been considered as a limit-state of its 3D counterpart \citep{Moffatt1967on,Sommeria1982why,Shats2010turbulence}, perhaps because both very often coexist \citep{Celani2010turbulence}. For example, atmospheric flows are quasi-2D at large, continental scales, but 3D nearer to topographic scales \citep{Lindborg1999atmospheric}. A similar spectral split exists in magnetohydrodynamic turbulence \citep{Baker2018inverse} and in turbulence in thin channels \citep{Benavides2017critical}. Quasi-2D turbulence appears progressively rather than through a bifurcation, and is controlled by the constraint driving \twodity\ \citep{Sommeria1982why,Potherat2014why,Benavides2017critical}. Turbulent transition in quasi-2D shear flows differs from the switch between 3D and 2D turbulence as it is expected to arise suddenly, out of (quasi-)2D finite amplitude instabilities. Since 3D mechanisms are excluded, the question is whether there exists a quasi-2D subcritical transition pathway from the laminar to the turbulent state. This is particularly crucial in shear flows as the subcritical lift-up transition is progressively suppressed as \twodity\ is established \citep{Cassels2019from3D}. The key result presented in this paper is the discovery of a transition from the quasi-2D laminar state to subcritical quasi-2D turbulence in shear flows, and that the lack of a 3D bypass mechanism gives way to an alternative nonlinear 2D mechanism which, unlike its 3D counterpart, relies on \TS\ waves.
\begin{figure}
\centering
\includegraphics[width=0.55\textwidth]{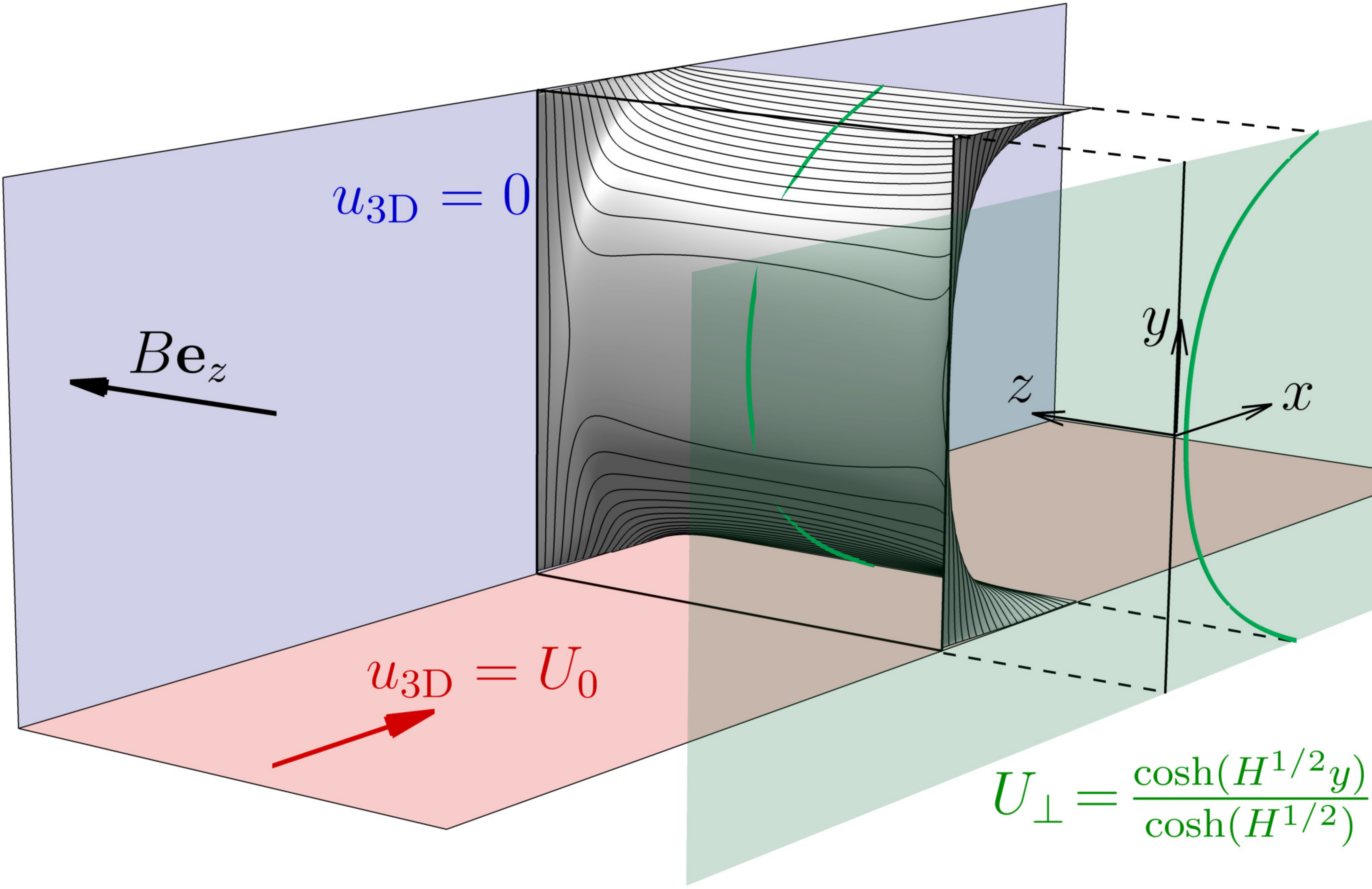}\\
	\caption{An example of a \qtwod\ flow is shown, namely a lateral wall-driven laminar duct flow under a transverse magnetic field. Here the solution (green profile; $H=10$) obtained from the \qtwod\ equations (\ref{eq:Q2Dc}) precisely approximates the $z$-average of the 3D streamwise velocity profile (grey isosurface; black contour lines) satisfying the quasi-static magnetohydrodynamic equations \citep{Muller2001magnetofluiddynamics}. Beyond this example, the results derived in this paper apply to a much wider variety of flows.}
    \label{fig:schematic}
\end{figure}

\section{Physical model}
\label{sec:phys-model}
All calculations are performed on a rectangular incompressible duct flow, with walls in the $(x,z)$ plane moving at a constant velocity $U_0\mathbf{e}_x$. The base flow is streamwise invariant and periodic boundary conditions are imposed in the streamwise direction ($x$).
The flow is assumed quasi-2D, \ie\ all quantities are invariant in $z$ except in thin boundary layers near the fixed walls in $(x,y)$ planes. Such flows are well described by the 2D, $z$-averaged \NavSto\ equations for $z$-averaged velocity and pressure supplemented by a linear friction term accounting for the friction these layers impart. With length, velocity, time and pressure respectively scaled by $L$, $U_0$, $L/U_0$ and $\rho U_0^2$, these equations are
\begin{equation}\label{eq:Q2Dc}
	\bnabp\bcdot\uvp = 0,\quad\p_t\uvp + \left(\uvp\bcdot\bnabp\right)\uvp = -\bnabp\pp + \Rey^{-1}\left(\bnp^2\uvp - H\uvp\right),
\end{equation}
with dimensionless boundary conditions $\uvp(y=\pm1) = (1,0)$,
where $\uvp=(\up,\vp)$, $\bnabp = (\p_x,\p_y)$, $\bnp^2=\p_x^2+\p_y^2$, $\rho$ is the fluid density, and $2L$ the distance between the moving walls. Friction parameter $H$ may be defined as appropriate to describe systems including duct flows under a strong transverse magnetic field $B\mathbf{e}_z$ \citep{Sommeria1982why}, thin films \citep{Buhler1996instabilities}, or flows with background rotation (with the addition of the Coriolis force) \citep{Pedlosky1987geophysical}.

We investigate perturbations $\uvph$ about the base flow $\Uvp(y) = (\cosh[H^{1/2}y]/\allowbreak\cosh[H^{1/2}],0)$ at $H=10$. Here linear perturbations become unstable at $\ReyCrit = 79123.2$. At $\rrc = 0.9$, the maximum linear transient growth occurs at a streamwise wavenumber $\alphaOpt = 1.49$, whereas the minimum exponential decay occurs at $\amax = 0.979651$.
The linearly optimised initial perturbation maximizing growth in the functional $G=\lVert\uvph(t=\tau)\rVert/\lVert\uvph(t=0)\rVert$ is sought following \citet{Barkley2008direct} for a prescribed target time $\tau$ and wavenumber $\alpha$. $G$ represents the gain in perturbation kinetic energy under the norm $\lVert\uvph\rVert = \int \uvph \bcdot \uvph \,\mathrm{d}\Omega$, over computational domain $\Omega$. Optimization is performed on the linearised rather than the full nonlinear equation, though both return practically identical results for this problem \citep{Camobreco2020role}. The choice of $\alpha$ is based on the decay rate of the leading direct eigenmode, obtained by decomposing perturbations into normal modes $(\mathbf{\tilde{u}_\perp}, \tilde{p}_\perp) \exp(\ii[\alpha x-\omega t])$ of complex frequency $\omega$. A discretised direct eigenvalue problem $-\ii\omega\vtp = \bm{L}\vtp$ is solved in MATLAB via \texttt{eigs}($\ii\bm{L}$). The linear evolution operator $\bm{L}$ is constructed following \citet{Trefethen1993hydrodynamic,Schmid2001stability}. The discretised adjoint eigenvalue problem $\ii\omega^\ddag\xtp = \bm{L}^\ddag\xtp$ is also considered, where the linear adjoint operator $\bm{L}^\ddag$ is derived following \citet{Schmid2001stability}.

To support the classification of initial conditions realizing turbulence, streamwise Fourier spectra of kinetic energy are computed at selected instants in time at $21$ equi-spaced $y$-values spanning the channel. At each $y$-location, a Fourier transform is obtained along $x$, with coefficients $c_\kappa(y) = \Nf^{-1}\sum_{n=0}^{\Nf-1}\lvert\uvph(x_n,y)\rvert^2\mathrm{exp}(-2\pi \ii \kappa n/\Nf)$,
where $x_n=2\pi n/\alpha\Nf$ spans the streamwise-periodic domain. Here for convenience the coefficient $\Nf^{-1}$ is applied to the forward transform rather than its inverse. Instantaneous mean Fourier coefficients $\meanfoco$ are then obtained by averaging $\lvert c_\kappa\rvert$ over $y$.

Time evolution of the full Q2D equations or the linear forward and adjoint systems is computed numerically using a primitive variable spectral element solver \citep{Hussam2012optimal, Cassels2019from3D, Camobreco2020role, Camobreco2020transition}. The $x$-$y$ plane is discretised with quadrilateral elements (12 by 48) featuring polynomial basis functions of order $\Np=19$ \citep{Camobreco2020role,Camobreco2020transition,Camobreco2021stability}. Time integration is via third-order backward differencing, with operator splitting \citep{Karniadakis1991high}. The time step size is initially set to $\Delta t = 1.25\times10^{-3}$, and is reduced once turbulence emerges to maintain stability. The initial condition is composed of the laminar base flow and a perturbation computed via linear transient growth optimization. The domain length is matched to the wavelength minimizing the decay rate of the leading eigenmode. The perturbation amplitude is normalised to the energy $\Ezero$ required for each simulation.

\section{Evidence of subcritical turbulence}
\label{sec:evidence-subcrit-turb}
The starting point is to determine whether subcritical turbulence in quasi-2D shear flows exists (for $\Rey<\ReyCrit$). In the subcritical regime, turbulence originates from perturbations that are sufficiently intense to activate nonlinear amplification mechanisms that infinitesimal ones cannot. Unlike 3D flows, there is evidence that seeding the subcritical laminar shear flow with even high levels of noise does not ignite turbulence \citep{Camobreco2020transition}. Seeking turbulence, but not necessarily its most efficient trigger, the laminar state is seeded with optimal transient growth perturbations of different energies, and evolved until the flow either returns to its initial laminar state or becomes turbulent.
\begin{figure}
\centering
\parbox{0.5\textwidth}{\small{(a)}}\parbox{0.5\textwidth}{\small{(b)}}\\
\parbox{0.025\textwidth}{\rotatebox{90}{\hspace{12pt}\small{$\Euv$}}}\parbox{0.475\textwidth}{\includegraphics[width=0.475\textwidth,trim={0 0.03cm 0 0},clip]{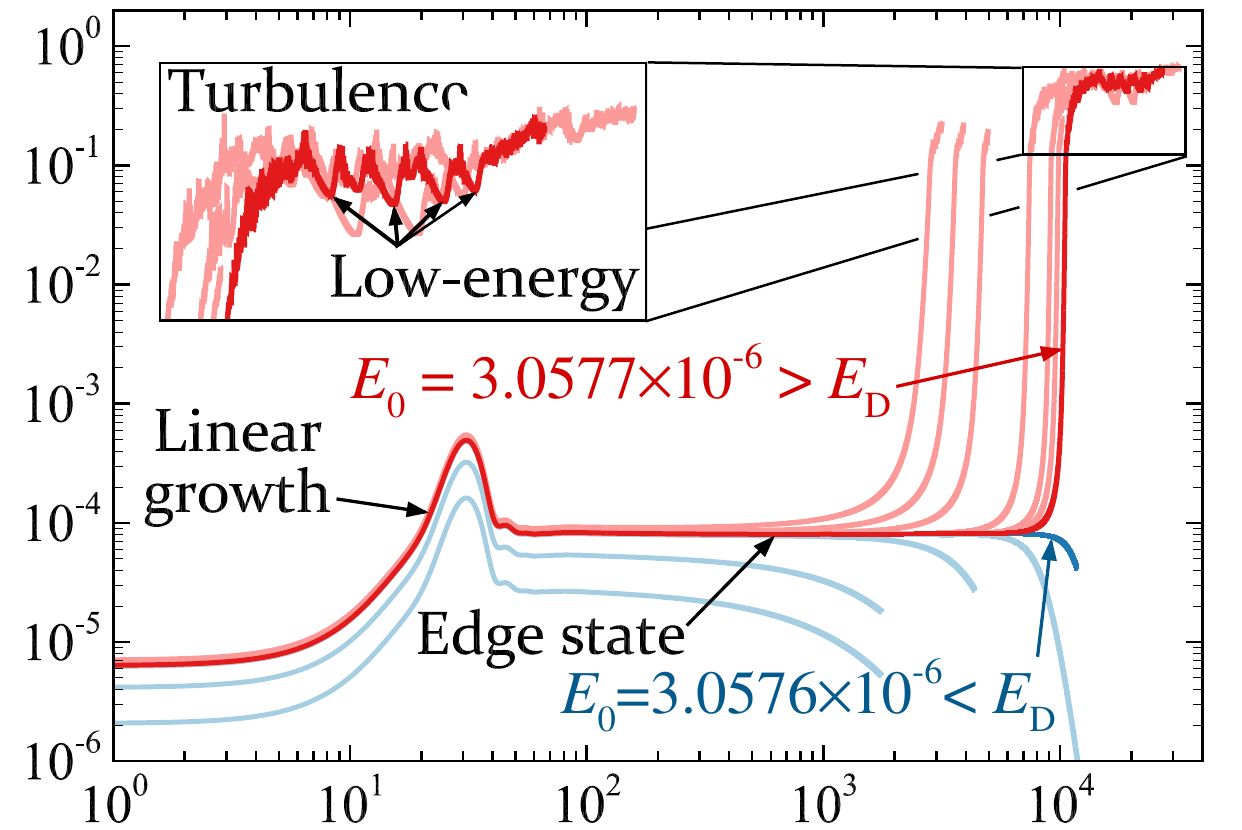}\\
 \centering $t$}%
\parbox{0.025\textwidth}{\rotatebox{90}{\hspace{12pt}\small{$\meanfoco$}}}\parbox{0.475\textwidth}{\includegraphics[width=0.475\textwidth,trim={0 0.03cm 0 0},clip]{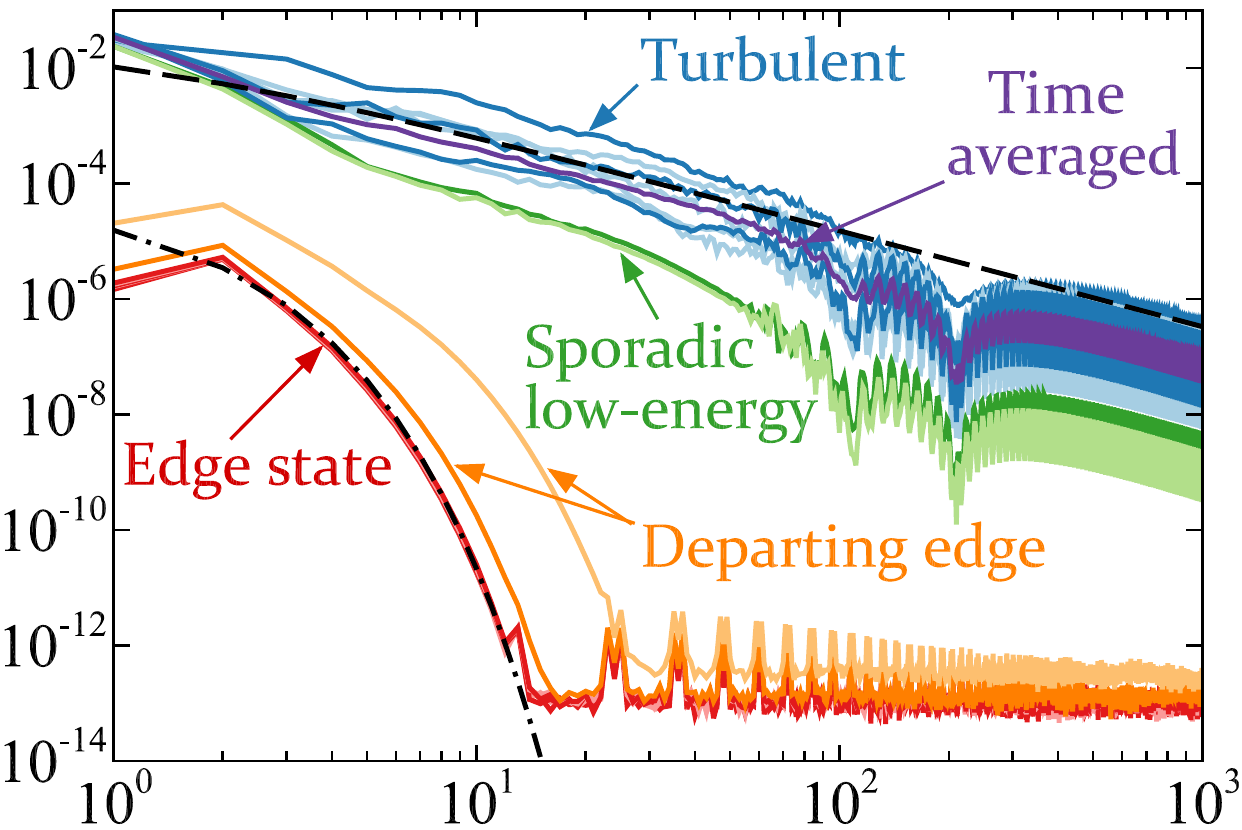}\\
 \centering $\kappa$}
\parbox{0.5\textwidth}{\small{(c)}}\parbox{0.5\textwidth}{\small{(d)}}\\
\parbox{0.5\textwidth}{\parbox[t]{0.5\textwidth}{\centering\includegraphics[width=0.475\textwidth]{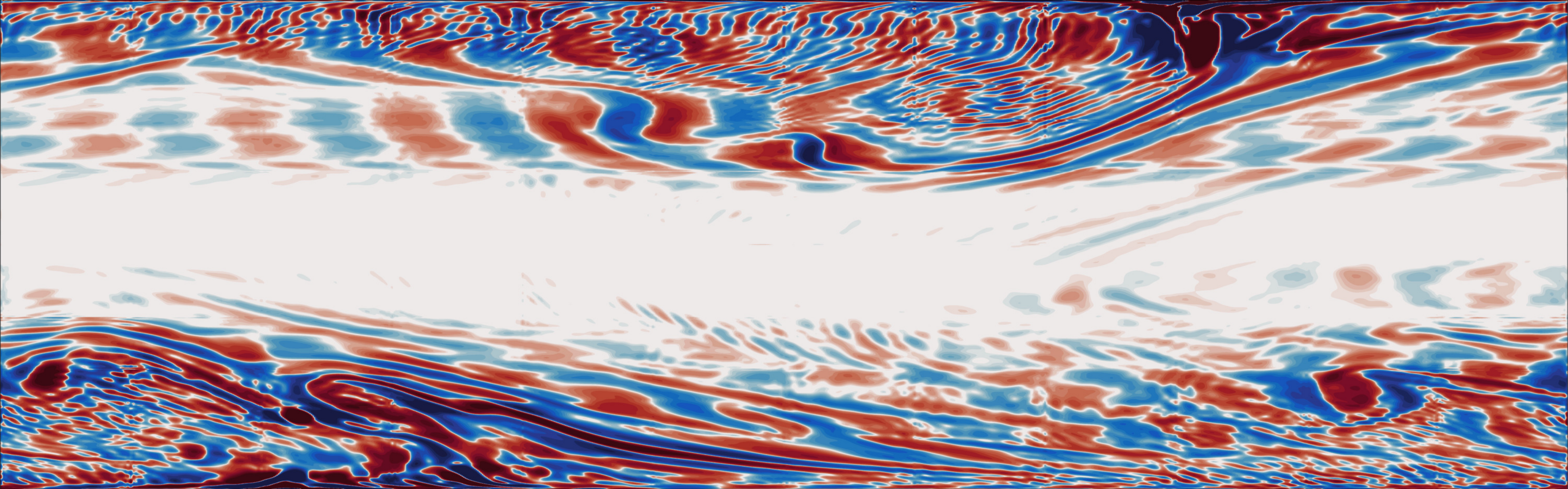}}\\[2pt]
\centering\small{$\sign(\hat{\omega}_{z,\lvert\kappa\rvert\geq 10})\lvert\hat{\omega}_{z,\lvert\kappa\rvert \geq 10}\rvert$} \\[2pt]
\centering\includegraphics[width=0.475\textwidth]{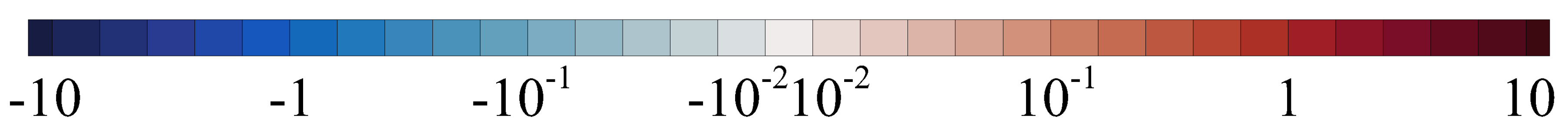}}%
\parbox{0.025\textwidth}{\rotatebox{90}{\hspace{12pt}\small{$\mathrm{d}(\log A)/\mathrm{d}t$}}}%
\parbox{0.475\textwidth}{\includegraphics[width=0.475\textwidth,trim={0 0.03cm 0 0},clip]{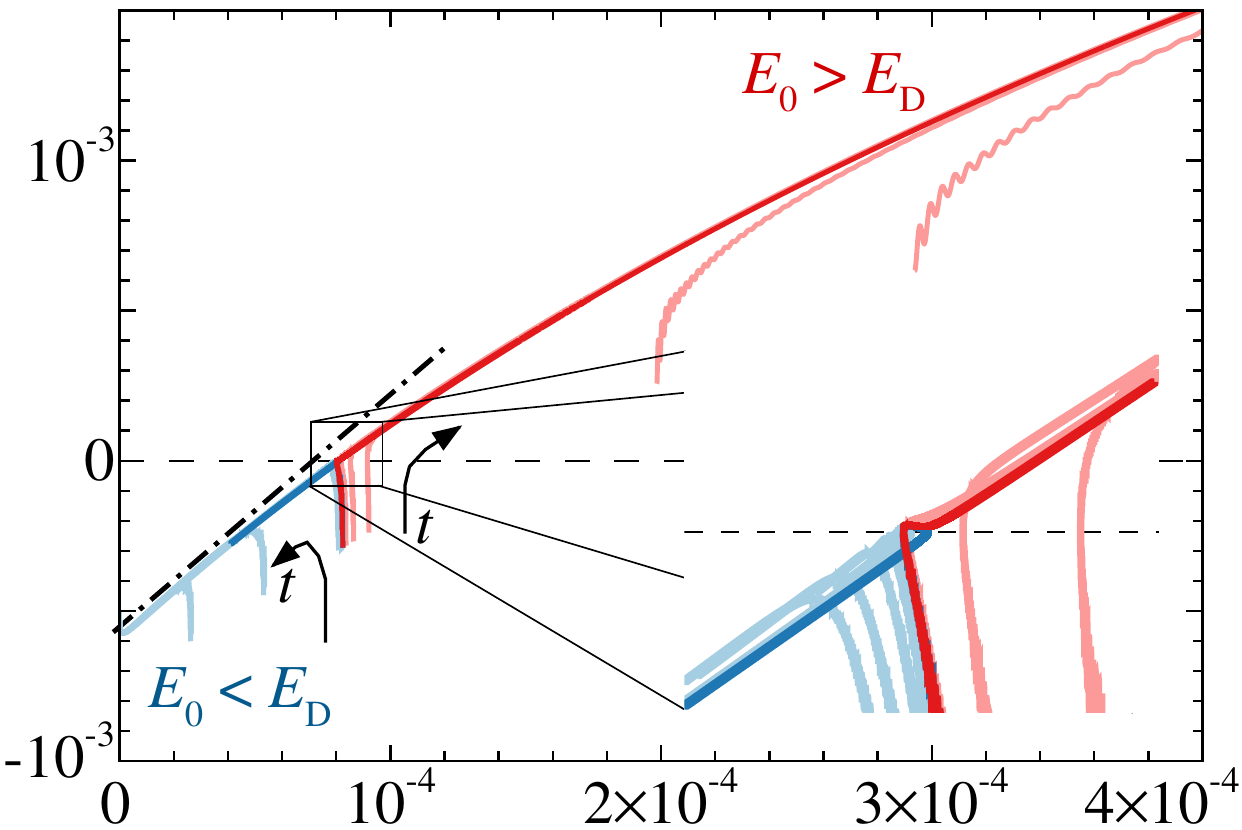}\\
 \centering $A^2$}
\caption{(a) Kinetic energy time history showing the nonlinear evolution of linear optimals at $\rrc = 0.9$, $\amax$: when $\Ezero<\ELD$ (blue) the flow visits the edge but relaminarises, whereas it becomes turbulent for $\Ezero>\ELD$ (red). (b) Fourier spectra at select instants in time at $\Ezero=3.0577\times10^{-6}>\ELD$. Dash-dotted line, $\exp(-3\kappa/2)$ trend. Dashed line, $\kappa^{-5/3}$ trend. (c) Streamwise high-pass filtered snapshot of spanwise vorticity $\lvert\hat{\omega}_{z,\lvert\kappa\rvert\geq 10}\rvert$ from DNS at $\rrc = 0.9$, $\amax$, indicating quasi-2D turbulence at $t=1.1\times10^4$. (d) Data from (a) re-plotted under the framework of the Stuart--Landau model. The instantaneous growth rate of the perturbation amplitude $A$ is plotted against $A^2$. The data exhibits a collapse onto a common curve exhibiting the signature of a subcritical bifurcation, \ie\ approaching the eigenmode growth rate $\mathrm{Im}(\omega)=-5.94084\times 10^{-4}$ as $A\rightarrow 0^+$ with a positive gradient. For guidance, the dot-dashed line is tangent to this curve at $A=0$.}
\label{fig:delin_ex}
\end{figure}
\Fig~\ref{fig:delin_ex}(\emph{a}) depicts a representative set of the aforementioned simulations at a Reynolds number $\rrc=0.9$. A turbulent state is reached for any normalised initial energy $E_0>\ELD$, where $E_0=E/\EB$. $E$ is the kinetic energy of the disturbance, and $\EB$ is the energy of the laminar base flow. The delineation energy $\ELD$, found when seeding the flow with optimal perturbations from linear transient growth analysis, lies within $3.0576\times10^{-6}<\ELD<3.0577\times10^{-6}$. Evidence of a turbulent state is found in the energy spectra $\meanfoco(\kappa)$ of \fig~\ref{fig:delin_ex}(\emph{b}): while low energy states contain energy above the noise floor (around $10^{-13}$) in only a few of the lower wavenumber modes, all modes are energised in the turbulent cases \citep{Grossmann2000onset}, with an extended inertial range following $\meanfoco(\kappa)\sim \kappa^{-5/3}$ \citep{Tabeling2002twodim-turb-physicist-appr}. These features are characteristic of turbulence, a snapshot of which is visualised in \fig~\ref{fig:delin_ex}(\emph{c}). This visualisation employs a streamwise high-pass filter to remove the otherwise occluding large-scale TS wave structures. This filter reveals smaller-scale structures being entrained from the side-wall boundary layers into the channel interior. Instances of this are visible near the bottom wall at the upstream end of the domain, and further downstream near the top wall. The respective locations of these features align with the entrainment regions of the underlying \TS\ wave.

With the existence of subcritical turbulence in quasi-2D flows now established, two remarkable features emerge: First, the turbulence is intermittent in time, exhibiting sporadic regressions to a low energy state differing from the original laminar state (\fig~\ref{fig:delin_ex}\emph{a}). This is reminiscent of the spatial intermittency in various 3D shear flows \citep{Cros2002spatiotemporal, Barkley2007mean, Moxey2010distinct, Brethouwer2012turbulent, Khapko2014complexity}. Second, transition to indefinitely sustained turbulence was only found for $\Rey \gtrsim 0.8\ReyCrit$. Hence the Reynolds numbers required to sustain turbulence are much higher than in 3D flows. Over $0.4\lesssim \rrc \lesssim 0.8$, only a single turbulent episode was observed, with finite lifetime proportional to $\Rey$.

Verification that the turbulence reported herein originates from a subcritical instability is determined using the Stuart--Landau model \citep{Drazin2004hydrodyamic} following the approach detailed in \cite{Sapardi2017linear}, a brief outline of which is explained here. The time history of a measure of the disturbance amplitude $A(t) = \sqrt{\int_\Omega \lvert\uvph\rvert^2\,\mathrm{d}\Omega}$ is taken, with subcritical bifurcation evolution characterised by an increase in $\mathrm{d}(\log A)/\mathrm{d}t$ with increasing amplitude near $A=0$. Beyond the critical Reynolds number, an infinitesimal disturbance achieves super-exponential growth before saturating. Below the critical Reynolds number, small disturbances decay, while larger-amplitude disturbances grow. This behaviour is observed in \fig~\ref{fig:delin_ex}(\emph{d}). Cases bracketing $\ELD$ approach a common curve having the expected subcritical profile; cases with $E<\ELD$ then decay ($\mathrm{d}(\log A)/\mathrm{d}t<0$, $A\rightarrow 0$), while $E>\ELD$ cases grow towards turbulence.

\section{Nonlinear Tollmien--Schlichting waves are the tipping point between laminar and turbulent states}
\label{sec:TS-tipping-point}
\begin{figure}
\centering
\parbox{0.8\textwidth}{%
\raggedright\small{(a)}\\
\centering\hspace{0.025\textwidth}\includegraphics[width=0.45\columnwidth]{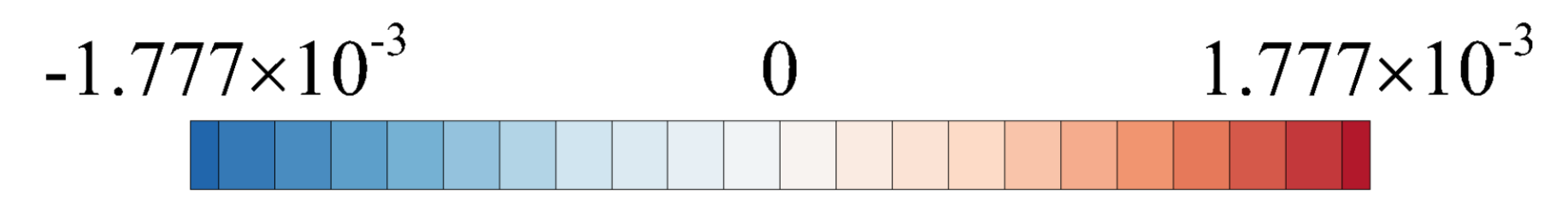}\\
\parbox{0.025\textwidth}{\rotatebox{90}{\hspace{14pt}\small{$y$}}}\parbox{0.775\textwidth}{\includegraphics[width=0.775\textwidth]{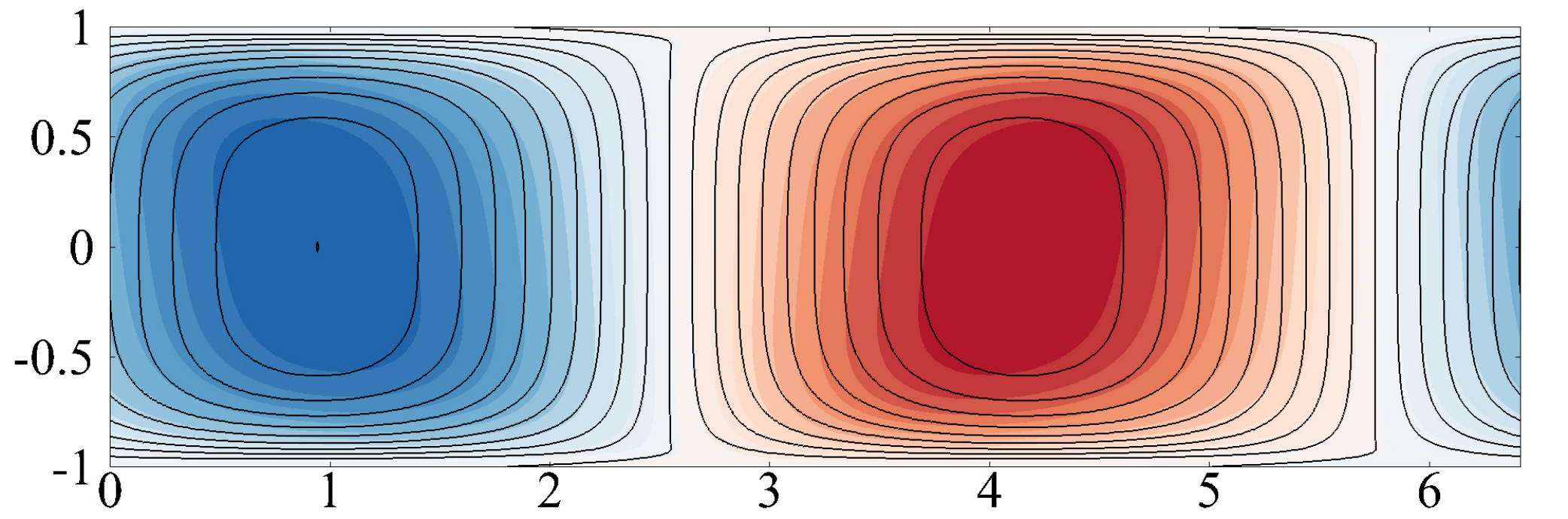}\\
 \centering $x$}\\
\raggedright\small{(b)}\\
\centering\parbox{0.025\textwidth}{\rotatebox{90}{\small{$y$}}}%
\parbox{0.775\textwidth}{\centering\hspace{0.025\textwidth}\color{mybd}{$\uot$}\hspace{40mm}\color{mygd}{$\wtt$} \\
\includegraphics[width=0.775\columnwidth]{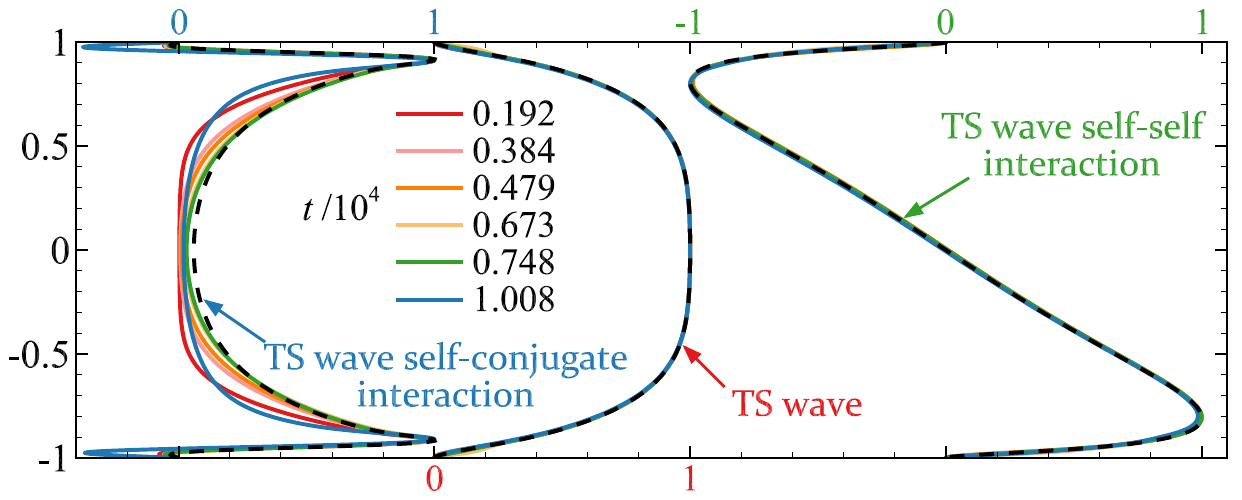}\\
\centering\color{myrd}{$\woo$}}}
\caption{(a) Flooded contours of spanwise velocity $\vhp$ from DNS at $\rrc = 0.9$, $\amax$, representing the edge state at $t=7.48\times10^3$ overlaid with contour lines showing the linear \TS\ wave $\woo$. Both sets of contours are equi-spaced between the minimum and maximum values of the respective fields. (b) Weakly nonlinear modes $\uot$, $\woo$, $\wtt$ (dashed black lines) and corresponding Fourier components from DNS at different times (colored lines) at $\kappa=0$ (streamwise), $1$ (spanwise) and $2$ (spanwise) with $\Ezero = 3.0577\times10^{-6}>\ELD$. The solution departs the edge at $t \approx 9\times10^3$.}
\label{fig:comp_DNS_WNL}
\end{figure}

Having established that subcritical turbulence exists, we now consider the pathway from the laminar base flow to the turbulent state. We seek the `edge state'; the tipping point from which the flow can either revert to its original laminar state, or become turbulent \citep{Skufca2006Edge}.
The edge state is reached for the initial perturbation \emph{delineation energy} $\ELD$, separating perturbations triggering turbulence from those decaying. As the initial energy approaches $\ELD$, the edge state persist for a longer duration. $\ELD$ is found iteratively with a bisection method \citep{Itano2001dynamics}. The edge state from \fig~\ref{fig:delin_ex}(\emph{a}) is visualised in \fig~\ref{fig:comp_DNS_WNL}(\emph{a}). It consists of a travelling wave of very similar topology to the infinitesimal \TS\ wave, suggesting they may play a role in the quasi-2D transition to turbulence.

To investigate this possibility, we calculated a weakly nonlinear flow state in which nonlinearities only arise out of combinations of the leading \TS\ wave. The weakly nonlinear equations are a more precise version of the perturbation equations compared to the linearised version used to calculate the leading eigenmode and the perturbation maximising transient growth. They are obtained by approximating the perturbation by the leading eigenmode, and truncating its governing equations to the third order in its amplitude. For example, for a leading eigenmode of amplitude $\epsilon$, the $n^\mathrm{th}$ harmonic of the spanwise velocity component is written as
\begin{equation}\label{eq:amp_exp}
	\vhpn = \sum_{m=0}^{\infty}\epsilon^{\lvert n\rvert+2m}\tilde{A}^{\lvert n\rvert}\lvert\tilde{A}\rvert^{2m}\vhpnnm,
\end{equation}
where $\vhpnnm$ denotes a perturbation ($n$ refers to the harmonic, $\lvert n\rvert+2m$ to the amplitude order) and $\tilde{A} = A/\epsilon$ is the normalised amplitude. Nonlinear self-interaction of the linear mode $\woo$ excites a second harmonic $\wtt$, which is compared to the $\kappa=2$ harmonic from DNS. Nonlinear interaction between the linear mode $\woo$ and its complex conjugate $\wno$ generates a modification to the base flow $\uot$, which is compared to the $\kappa=0$ harmonic from DNS. The full equations governing the $n^\mathrm{th}$ harmonic of the base flow and perturbation follow from insertion of this decomposition into Eq.~(\ref{eq:Q2Dc}); they are expressed in full in \citet{Camobreco2020transition} and the general method is detailed in \citet{Hagan2013weakly}. The full nonlinear evolution of the flow is obtained by solving the system (\ref{eq:Q2Dc}) numerically.

This technique facilitates a comparison between the fully nonlinear flow state (with all possible modes) obtained from DNS, with its asymptotic approximation to second order in the perturbation amplitude. The streamwise $\kappa=1$ Fourier mode extracted from DNS is directly compared to the linearly computed modal instability in \fig~\ref{fig:comp_DNS_WNL}(\emph{a}), showing close agreement.

\Fig~\ref{fig:comp_DNS_WNL}(\emph{b}) compares velocity profiles from each of the modes accounted for in the weakly nonlinear analysis with the corresponding Fourier components of the same wavelength extracted from the full DNS evolved from the linear optimal with $\alpha=\amax$ with $\Ezero$ close to $\ELD$. Both the leading eigenmode ($\kappa=1$) and its nonlinear interaction ($\kappa=2$) matched their DNS counterpart to high precision in the early stage of evolution. The modulated streamwise-independent ($\kappa=0$) component exhibited small differences at this stage, that vanished as the influence of our particular choice of initial condition did. Additionally, the cumulated kinetic energy of all three components forming the weakly nonlinear approximation represents over $93.7$\% %to $93.9$\%
of the total energy in the DNS while on the edge. This proves that the build-up of the edge state originates almost exclusively from the dynamics of the \TS\ wave. As such, this transition mechanism differs radically from its counterpart in 3D flows \citep{Zammert2019transition}, where a bypass transition involving rapidly growing streamwise structures takes place at such low Reynolds numbers that \TS\ waves are too damped to contribute.

With the pathway to the edge state and then to turbulence now clarified, the question remains as to its robustness against the choice of initial condition. Thus, DNS were performed with the initial conditions chosen as the modes optimizing growth at increasingly large times $T = \tau/\tauOpt$ (where $\tauOpt$ is the time of optimal growth for a given
wavenumber $\alpha$, \ie\ $\tauOpt\simeq31.0$ for $\alpha=\alpha_{\rm \max}$ at $\rrc=0.9$).
The corresponding delineation energy provides a measure of how easily these transients ignite turbulence and therefore of their role in doing so.
\begin{figure}
\centering
\parbox[t][][t]{0.475\textwidth}{%
\raggedright\small{(a)}\\
\parbox[t][][t]{0.025\textwidth}{\rotatebox{90}{\small{$y$}}}%
\parbox{0.45\textwidth}{\includegraphics[width=0.45\textwidth]{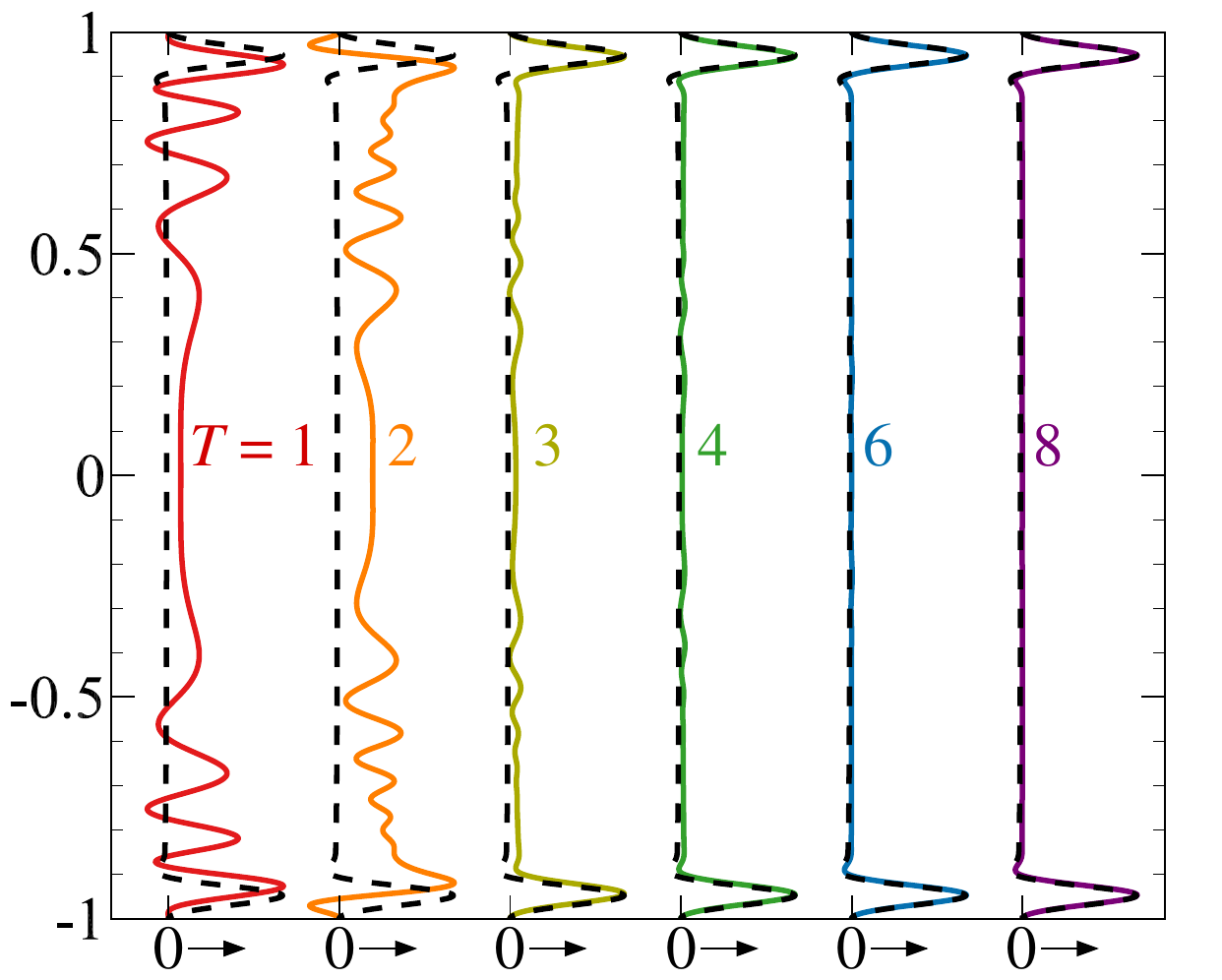}}
\center\hspace{0.025\textwidth}\small{$\lvert\vtp\rvert$}}%
\parbox[t][][t]{0.5\textwidth}{%
\raggedright\small{(b)}\\
\centering\parbox{0.025\textwidth}{\rotatebox{90}{\hspace{6pt}\small{\color{mybd}{$E(T)$}, \color{mybd}{$\max(E(t))$}}}}%
\parbox{0.45\textwidth}{\includegraphics[width=0.45\textwidth]{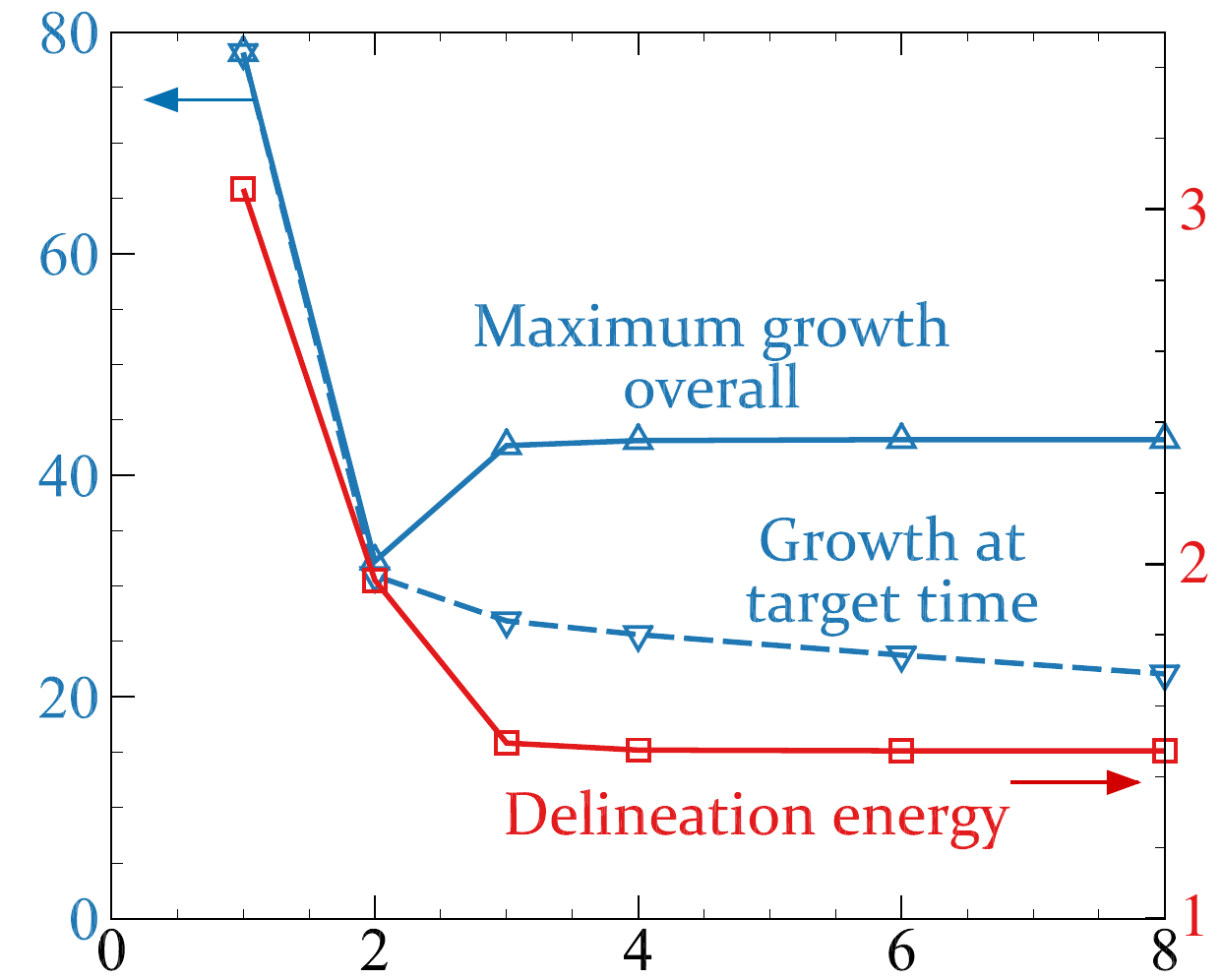}}%
\parbox{0.025\textwidth}{\rotatebox{90}{\hspace{6pt}\small{\color{myrd}{$10^6\ELD$}}}}\\
\small{$T = \tau/\tauOpt$}}
\caption{(a) Comparison between linear transient growth initial conditions with various $\tau=T\tauOpt$ (solid lines) and the leading adjoint mode (dashed lines), horizontally shifted for clarity.
(b) Delineation energies from DNS and energy growth ratios from linear analysis for each $T$.}
\label{fig:int_opt_comp}
\end{figure}

In these simulations, all optimals for a given $\alpha$ evolved into the same edge state. Further, as $T$ increases, the profiles of spanwise velocity of the initial condition optimizing growth with $\alpha=\amax$ converge toward the leading adjoint mode (\fig~\ref{fig:int_opt_comp}\emph{a}), with a corresponding monotonic reduction in the delineation energy with increasing $T$ (\fig~\ref{fig:int_opt_comp}\emph{b}). By $T=8$, the delineation energy of the optimal matches that of the leading adjoint to approximately $\pm 0.01\%$.
\Fig~\ref{fig:int_opt_comp}(\emph{b}) further shows that these initial conditions lead to reduced transient growth, yet are \emph{more} efficient at triggering turbulence: that is to say, the delineation energy $E_D$ decreased approximately twofold as the initial condition morphed from the linear optimal mode to the linear adjoint mode. This was found to be consistent when the streamwise wavenumber was varied. \cite{DuguetBrandtLarsson2010plane-couette} similarly showed that maximising transient growth does not necessarily imply an easier transition. Hence the leading adjoint eigenmode, which by construction optimally energises the \TS\ wave, is a more efficient initial condition to reach turbulence than any initial condition producing optimal transient growth. Consequently, optimal growth does not favour the transition, unlike in 3D shear flows where the transient growth associated with the lift-up mechanism is an essential element of the transition process \citep{Reddy1998stability,Pringle2012minimal}.

The same procedures applied at lower $\rrc$ produced the same reduction in delineation energy with increasing $T$ and exhibited edge states independent of $T$. For $0.3 \lesssim \rrc \lesssim 0.8$, after departing the edge, a secondary stable state formed \citep{Jimenez1990transition,Falkovich2018turbulence}, again independent of the initial condition.

\section{Discussion and concluding remarks}
\label{sec:disc-concl}
In conclusion, subcritical turbulence exists in quasi-2D shear flows and can be reached directly, rather than via an intermediate 3D state. The 2D transition mechanism bears important similarities with its 3D cousin: it is ignited by a perturbation of finite amplitude and first reaches an edge state that is seemingly independent of this initial perturbation. The edge state subsequently breaks down into a turbulent state if the initial perturbation energy exceeds the delineation energy $\ELD$ for that particular perturbation. As in the 3D problem, the attained turbulence is not yet fully developed \citep{Wygnanski1973p1,Wygnanski1975p2}. Departure from fully established turbulence in quasi-2D shear flows expresses as time intermittency, with sporadic retreats to a low energy state different from the base laminar flow.

Conversely, the subcritical transition in quasi-2D shear flows exhibits specificities that distinguish it sharply from the 3D one. Chiefly, the lift-up mechanism that underpins transitions in 3D can be ignited at criticality so low that the \TS\ waves are strongly suppressed by the linear dynamics, despite being the least damped infinitesimal perturbation. Thus, they are not observed in the 3D transition. In quasi-2D flows by contrast, the 3D mechanism is absent and our study demonstrates that the dynamics are dominated by the \TS\ waves, with the edge state resulting directly from their weakly nonlinear evolution. In a quasi-2D flow, the \TS\ wave instability directly connects the base flow to turbulence via a subcritical bifurcation, in stark contrast to a 3D flow in which the saddle-node bifurcation is disconnected from the base state \citep{Khapko2014complexity}. This may also explain why transition in quasi-2D flows is relatively weakly subcritical: at lower $\rrc$, \TS\ waves are so strongly linearly damped that their nonlinear growth is stifled.

This new transition mechanism reopens many questions resolved in the 3D case: How does the intermittency or localization of the turbulence evolve into the supercritical regime, \eg following the mechanism outlined by \citet{Mellibovsky2015mechanism}? Does the transition to the fully turbulent state obey a second-order phase transition of the universality class of directed percolations as for other shear flows \citep{Lemoult2016directed}? While the thermodynamic formalism used by \citet{WangLiE2015pnas} indicates that \twod\ Poiseuille flow loses stability in a manner consistent with a continuous phase transition, in the \qtwod\ case linear friction may impact the development and interaction of unstable travelling waves, and so this remains an open question. Separately, much remains to be discovered on the subcritical response to finite amplitude perturbations: how does the delineation energy vary with criticality, especially considering the relatively short subcritical range in which turbulence can be sustained? Can this subcritical response be manipulated to prevent or promote turbulence (for example, to enhance heat transfer in the heat exchangers of plasma fusion reactors)? These questions call for expensive numerical simulations, but also for experiments with well controlled perturbations, since this first evidence of subcritical transitions in quasi-2D shear flows is currently purely numerical.

While we have established a scenario for transition involving a purely 2D mechanism, 3D mechanisms could still compete with this scenario and trigger a subcritical 3D transition. Whether one of the other scenarios dominates cannot be determined on the basis of Eq.~(\ref{eq:Q2Dc}) only as the 3D mechanisms are specific to the physical process promoting the emergence of quasi-2D dynamics. As such, whether a purely quasi-2D subcritical transition to turbulence can be observed in practice remains to be determined in particular cases. The question could be addressed either with 3D numerical methods or experiments on MHD or rotating flows, or in Hele-Shaw cells, for example.

\acknowledgments
The authors thank Dr.\ Susanne Horn and Dr.\ Chris Pringle for their helpful feedback. C.J.C.\ was supported by the Australian Government Research Training Program (RTP). This research was supported by Australian Research Council Discovery Grant DP180102647 and Royal Society International Exchanges Grant IE170034. Computations were possible thanks to the National Computational Infrastructure (NCI), Pawsey Supercomputing Centre, and the Monash e-Research Centre.

\end{document}